\documentclass{birkmult}
 \newtheorem{thm}{Theorem}[section]

 \theoremstyle{definition}
 \newtheorem{defn}[thm]{Definition}
 \theoremstyle{remark}
 \newtheorem{rem}[thm]{Remark}
 \newtheorem*{ex}{Example}
 \numberwithin{equation}{section}

\usepackage{graphics, color, amsmath, amsfonts, amssymb, amsthm, amscd}
\usepackage{epsf}
\usepackage{psfrag}

\makeatletter
\def\captionfont@{\footnotesize}
\def\captionheadfont@{\scshape}
\long\def\@makecaption#1#2{%
  \vspace{2mm}
  \setbox\@tempboxa\vbox{\color@setgroup
    \advance\hsize-6pc\noindent
    \captionfont@\captionheadfont@#1\@xp\@ifnotempty\@xp
        {\@cdr#2\@nil}{.\captionfont@\upshape\enspace#2}%
    \unskip\kern-6pc\par
    \global\setbox\@ne\lastbox\color@endgroup}%
  \ifhbox\@ne 
    \setbox\@ne\hbox{\unhbox\@ne\unskip\unskip\unpenalty\unkern}%
  \fi
  \ifdim\wd\@tempboxa=\z@ 
    \setbox\@ne\hbox to\columnwidth{\hss\kern-6pc\box\@ne\hss}%
  \else 
    \setbox\@ne\vbox{\unvbox\@tempboxa\parskip\z@skip
        \noindent\unhbox\@ne\advance\hsize-6pc\par}%
\fi
  \ifnum\@tempcnta<64 
    \addvspace\abovecaptionskip
    \moveright 3pc\box\@ne
  \else 
    \moveright 3pc\box\@ne
\nobreak
\vskip\belowcaptionskip
\fi
\relax
}
\makeatother

\begin{document}
%
%
%
%
%
%
%
%
%
\title[Ising model fog drip]
 {Ising model fog drip: the first two droplets}
\author[Ioffe]{Dmitry Ioffe}

\address{%
Faculty of Industrial Engineering and Management,\\
Technion, 
Haifa 32000,\\
Israel}

\email{ieioffe@ie.technion.ac.il}

\author[Shlosman]{Senya Shlosman}
\address{
Centre de Physique Theorique, CNRS, \\
Luminy 
Case 907,\\
13288 Marseille, Cedex 9, France
}
\email{shlosman@cpt.univ-mrs.fr}
\subjclass{Primary: 82B20, 82B24 Secondary: 60G60  }

\keywords{Ising model, Phase segregation, Condensation}

\date{July 31, 2007}

\begin{abstract}
We present here a simple model describing coexistence of solid and vapour 
phases. The two phases are separated by an interface. We show that when the 
concentration of supersaturated vapour reaches the dew-point, the droplet of 
solid is created spontaneously on the interface, adding to it a monolayer of a 
\textquotedblleft visible\textquotedblright\ size. 
\end{abstract}

\maketitle

\section{Introduction: Condensation phenomenon in the Ising model} 
The phenomenon of droplet condensation in the framework of the Ising model was 
first described in the papers \cite{DS1}, \cite{DS2}. It deals with the 
following situation. Suppose we are looking at the Ising spins $\sigma_{t}%
=\pm1$ at low temperature $\beta^{-1}$, occupying a $d$-dimensional box 
$T_{N}^{d}$ of the linear size $2N$ with periodic boundary conditions. If we 
impose the canonical ensemble restriction, fixing the total mean 
magnetization, 
\[ 
\frac{M_{N}}{\left\vert T_{N}^{d}\right\vert }\,\overset{\Delta}{=}\,\frac 
{1}{\left\vert T_{N}^{d}\right\vert }\,\sum\sigma_{t}, 
\] 
to be equal to the spontaneous magnetization, $m^{\ast}\left(  \beta\right) 
>0$, then the typical configuration that we see will look as a configuration 
of the $\left(  +\right)  $-phase. That means that the spins are taking mainly 
the values $+1,$ while the values $-1$ are seen rarely, and the droplets of 
minuses in the box $T_{N}^{d}$ are at most of the size of $K\left(  d\right) 
\ln N.$ We want now to put more $-1$ particles into the box $T_{N}^{d},$ and 
we want to see how the above droplet picture would evolve. That means, we want 
to look at the model with a different canonical constraint: 
\[ 
M_{N}=m^{\ast}\left(  \beta\right)  \left\vert T_{N}^{d}\right\vert -b_{N}, 
\] 
$b_{N}>0.$ It turns out that if $b_{N}$-s are small, nothing is changed in the 
above picture; namely, if%
\[ 
\frac{b_{N}}{\left\vert T_{N}^{d}\right\vert ^{\frac{d}{d+1}}}\rightarrow 
0\text{ as }N\rightarrow\infty, 
\] 
then in the corresponding canonical ensemble all the droplets are still 
microscopic, not exceeding $K\left(  d\right)  \ln N$ in linear size. On the 
other hand, once 
\[ 
\liminf_{N\rightarrow\infty}\frac{b_{N}}{\left\vert T_{N}^{d}\right\vert 
^{\frac{d}{d+1}}}>0, 
\] 
the situation becomes very different: among many $\left(  -\right)  $-droplets 
there is one of the linear size $\ $of the order of $\left(  b_{N}\right) 
^{1/d}\geq N^{\frac{d}{d+1}},$ while all the rest of the droplets are still at 
most logarithmic. Therefore $b_{N}\sim\left\vert T_{N}^{d}\right\vert 
^{\frac{d}{d+1}}$ can be called the \textit{condensation threshold
, or 
dew-point
}. The behavior of the system \textit{at the threshold scale, } i.e. 
for $b_{N}=c\left\vert T_{N}^{d}\right\vert ^{\frac{d}{d+1}}\left( 
1+o_{N}\left(  1\right)  \right)  ,$ is considered in the 2D case in 
\cite{BCK}. Sharp description of the transition \textit{inside the threshold} 
is considered in \cite{HIK}. 
 
The above condensation picture suffers from one (largely esthetic) defect: 
both below and immediately above the condensation threshold the droplets are 
\textquotedblleft too small to be visible\textquotedblright, i.e. they are of 
the size sublinear with respect to the system size. This defect was to some 
degree bypassed in \cite{BSS}. It is argued there on heuristic level, that in 
the low-temperature 3D Ising model in the regime when $b_{N}$ is already of 
the volume order, i.e. $b_{N}\sim\nu N^{3},$ the sequence of condensations 
happens, with \textquotedblleft visible\textquotedblright\ results. In such 
regime one expects to find in the box $T_{N}^{3}$ a droplet $\Gamma$ of 
$\left(  -\right)  $-phase, of linear size of the order of $N,$ having the 
approximate shape of the Wulff crystal, which crystal at low temperatures has 
6 flat facets. One expects furthermore that the surface $\Gamma$ itself has 6 
flat facets, at least for some values of $b_{N}.$ However, when one further 
increases the \textquotedblleft supersaturation parameter\textquotedblright%
\ $b_{N},$ by an increment of the order of $N^{2},$ one expects to observe the 
condensation of extra $\left(  -\right)  $-particles on one of the flat facets 
of $\Gamma$ (randomly chosen), forming a monolayer $\mathfrak{m}$ of thickness 
of one lattice spacing, and of linear size to be $cN,$ with $c\geq 
c_{crit}=c_{crit}\left(  \beta\right)  ,$ with $c_{crit}N$ being smaller than 
the size of the facet. As $b_{N}$ increases further, the monolayer 
$\mathfrak{m}$ grows, until all the facet is covered by it. So one expects to 
see here the condensation of the supersaturated gas of $\left(  -\right) 
$-particles into a monolayer of linear size $\sim c_{crit}N,$ which is 
\textquotedblleft visible\textquotedblright. (Indeed, such monolayers were 
observed in the experiments of condensation of the Pb.) The rigorous results 
obtained in \cite{BSS} are much more modest: the model studied there is the 
Solid-on-Solid model, and even in such simplified setting the evidence of 
appearance of the monolayer $\mathfrak{m}$ of linear size is indirect. 
 
The purpose of the present paper is to consider another 3D lattice model, 
where one can completely control the picture and prove the above behavior to 
happen. Namely, we consider a system of ideal particles in the phase 
transition regime, and we put these phases -- the vapour phase and the solid 
phase -- into coexistence by applying the canonical constraint, i.e. by fixing 
the total number of particles. We study the interface $\Gamma,$ separating 
them, and we show that when we increase the total number of particles, 
the surface $\Gamma$ changes in the way described above. More precisely, we 
show that for some values of concentration the surface $\Gamma$ is essentially 
flat, but when the concentration increases up to the dew-point, a monolayer 
$\mathfrak{m}$ of a size at least $c_{crit}N$ appears on $\Gamma,$ with $N$ 
being the linear size of our system.   

\section{Informal description of the main result} 
 
In this section we describe our results informally. We will use the language 
of the Ising model, though below we treat rigorously a simpler model of the 
interface between two ideal particles phases. Ising model language makes the 
description easier; moreover, we believe that our picture holds for the Ising 
spins as well. 
 
Suppose we are looking at the Ising spins $\sigma_{t}=\pm1$ at low temperature 
$\beta^{-1}$ in a 3D box $B_{N}$ of the linear sizes $RN\times RN\times 2N.$ 
The parameter $R$ should be chosen sufficiently large in order to be 
compatible with the geometry of monolayer creation as described below.
 We impose $\left( 
+\right)  $-boundary conditions in the upper half-space $\left(  z>0\right) 
$, and $\left(  -\right)  $-boundary conditions in the lower half-space 
$\left(  z<0\right)  $. These $\left(  \pm\right)  $-boundary conditions force 
an interface $\Gamma$ between the $\left(  +\right)  $ and the $\left( 
-\right)  $ phases in $V_{N},$ and the main result of the paper \cite{D1} is a 
claim that the interface $\Gamma$ is rigid. It means that at any location, 
with probability going to 1 as the temperature $\beta^{-1}\rightarrow0,$ the 
interface $\Gamma$ coincides with the plane $z=0.$ If we impose the canonical 
ensemble restriction, fixing the total mean magnetization $M_{N}$ to be zero, 
then the properties of $\Gamma$ stay the same. 
 
We will now put more $-1$ particles into $V_{N};$ that is, we fix $M_{N}$ to 
be 
\[ 
M_{N}=-b_{N}=-\delta N^{2}, 
\] 
and we will describe the evolution of the surface $\Gamma$ as the parameter 
$\delta>0$ grows. The macroscopic image of this evolution is depicted
on Figure~1. 
 
\textbf{0. }$\mathbf{0\leq\delta<\delta^{1}}$ 
 
Nothing is changed in the above picture -- namely, the interface $\Gamma$ 
stays rigid. It is essentially flat at $z=0;$ the local fluctuations of 
$\Gamma$ are rare and do not exceed $K\ln N$ in linear size. 
 
\textbf{I. }$\mathbf{\delta^{1}<\delta<\delta^{2}}$ 
 
The monolayer $\mathfrak{m}_{1}$ appears on $\Gamma.$ This is a random 
outgrowth on $\Gamma,$ of height one. Inside $\mathfrak{m}_{1}$ the height of 
$\Gamma$ is typically $z=1,$ while outside it we have typically $z=0.$ 
 
For $\delta$ close to $\delta^{1}$ the shape of $\mathfrak{m}_{1}$ is 
\textit{the Wulff shape, }given by the Wulff construction, with the surface 
tension function $\tilde{\tau}^{2D}\left(  n\right)  ,$ $n\in\mathbb{S}^{1},$ 
given by 
\begin{equation} 
\tilde{\tau}\left(  n\right)  =\frac{d}{dn}\tau^{3D}\left(  m\right) 
\Bigm|_{m=\left(  0,0,1\right)  }. \label{11}%
\end{equation} 
Here $\tau^{3D}\left(  m\right)  ,$ $m\in\mathbb{S}^{2}$ is the surface 
tension function of the 3D Ising model, the derivatives in $\left( 
\ref{11}\right)  $ are taken at the point $\left(  0,0,1\right)  \in 
\mathbb{S}^{2}$ along all the tangents $n\in\mathbb{S}^{1}$ to the sphere 
$\mathbb{S}^{2}.$ The \textquotedblleft radius\textquotedblright\ of 
$\mathfrak{m}_{1}$ is of the order of $N,$ i.e. it equals to $r_{1}\left( 
\delta\right)  N,$ and as $\delta\searrow\delta^{1}$ we have $r_{1}\left( 
\delta\right)  \searrow r_{cr}>0.$ In particular, we never see a monolayer 
$\mathfrak{m}$ of radius smaller than $r_{cr}N.$ 
 
As we explain below $r_{cr}$ should scale like $R^{2/3}$. In particular, 
 it is
possible to choose $R$ in such a fashion that $R > 2 r_{cr}$ or, in other 
words, for values of $R$ sufficiently large the critical droplet fits into
 $B_N$. 

As $\delta$ increases, the monolayer $\mathfrak{m}_{1}$ grows in size, and at 
a certain moment $\delta=\delta^{\mathbf{1.5}}$ it touches 
the faces  of the box $B_N$ 
After that moment the shape of $\mathfrak{m}_{1}$ is different from the Wulff 
shape. Namely, it is \textit{the Wulff plaquette}\textit{ }(see \cite{SchS}), 
made from four segments on the four sides of the $RN\times RN$ square, 
connected together by the four quarters of the Wulff shape of radius 
$\tilde{r}_{1}\left(  \delta\right)  N.$ We have evidently $\tilde{r}%
_{1}\left(  \delta^{\mathbf{1.5}}\right)  =R/2 .$ 
As $\delta\nearrow\delta^{\mathbf{2}},$ the 
radius $\tilde{r}_{1}\left(  \delta\right)  $ decreases to some value 
$\tilde{r}_{1}\left(  \delta^{\mathbf{2}}\right)  N,$ with 
$\tilde{r}^{1}\left( 
\delta^{\mathbf{2}}\right)  >0.$ 
 
\textbf{II.} $\mathbf{\delta^{2}}\mathbf{<\delta<\delta^{2.5}}$ 
 
The second monolayer $\mathfrak{m}_{2}$ is formed on the top of $\mathfrak{m}%
_{1}.$ Asymptotically it is of Wulff shape with the 
radius $r_{2}\left(  \delta\right)  N,$ with $r_{2}\left( 
\delta\right)  
\searrow r_{2}^+\left(  \delta^{\mathbf{2}}\right)  $ as $\delta 
\searrow\delta^{2},$ with $r_{2}^+\left(  \delta^{2}\right)  >0.$
The first monolayer $\mathfrak{m}_{1}$ has a shape 
of Wulff plaquette with radius $\tilde{r}_{1}\left(  \delta\right)  ,$ which 
satisfies%
\[ 
\tilde{r}_{1}\left(  \delta\right)  =r_{2}\left(  \delta\right)  . 
\] 
A somewhat curious relation is:%
\[ 
r_{2}^+\left(  \delta^{\mathbf{2}}\right)  \text{ is strictly bigger 
than }\tilde{r}%
_{1}\left(  \delta^{\mathbf{2}}\right)  . 
\] 
In other words, the Wulff-plaquette-shaped monolayer $\mathfrak{m}_{1}$ 
undergoes a jump in its size and shape as the supersaturation parameter 
$\delta$ crosses the value $\delta^{2}.$ In fact, 
the monolayer $\mathfrak{m}_{1}$ 
shrinks in size: the radius $\tilde{r}_{1}\left( 
\delta\right)  $ increases as $\delta$ grows past $\delta^{\mathbf{2}}.$ 
 
\textbf{II.5} $\mathbf{\delta^{2.5}}\mathbf{<\delta<\delta^{3}}$ 

At the value $\delta=\delta^{\mathbf{2.5}}$ 
the growing monolayer $\mathfrak{m}_{2}$ 
meets the shrinking monolayer $\mathfrak{m}_{1},$ i.e. $r_{2}\left( 
\delta^{\mathbf{2.5}}\right)  =
\tilde{r}_{1}\left(  \delta^{\mathbf{2.5}}\right)  =R/2.$ Past the 
value $\delta^{\mathbf{2.5}}$ the two monolayers $\mathfrak{m}_{2}\subset 
\mathfrak{m}_{1}$ are in fact asymptotically equal, both having the shape of 
the Wulff plaquette with the same radius $\tilde{r}_{1}\left(  \delta\right) 
=\tilde{r}_{2}\left(  \delta\right)  ,$ decreasing to the value $\tilde{r}%
_{1}\left(  \delta^{\mathbf{3}}\right)  =\tilde{r}_{2}\left(  \delta 
^{\mathbf{3}}\right)  $ as $\delta$ increases up to $\delta^{\mathbf{3}}$. 
 
\textbf{III.} $\mathbf{\delta^{3}}\mathbf{<\delta<\delta^{4}}$ 
 
The third monolayer $\mathfrak{m}_{3}$ is formed, of the 
asymptotic radius $r_{3}\left( 
\delta\right)  N,$ with $r_{3}\left(  \delta\right)  \searrow r^+_{3}\left( 
\delta^{\mathbf{3}}\right)  $ as $\delta\searrow\delta^{\mathbf{3}},$ 
with $r^+_{3}\left( 
\delta^{\mathbf{3}}\right)  >0.$ The 
radii of two bottom Wulff plaquettes  $\tilde{r}_{1}\left(  \delta\right) 
=\tilde{r}_{2}\left(  \delta\right) = r_{3}\left(  \delta\right) $ 
decrease to the value ${r}%
_{3}^{+}\left(  \delta^{\mathbf{3}}\right)$
as $\delta$ decreases down to $\delta^\mathbf{3}{},$ with ${r}%
_{3}^{+}\left(  \delta^{\mathbf{3}}\right)  >\tilde{r}_{i}\left(  \delta 
^\mathbf{3}{}\right)  ,$ so 
the two Wulff plaquettes $\mathfrak{m}_{1},\mathfrak{m}%
_{2}$ shrink, jumping to a smaller area, as $\delta$ passes the threshold 
value $\delta^{\mathbf{3}}.$ 
 
\smallskip

... 

\smallskip

\begin{figure}[htb] 
\psfrag{a}{$\delta^{\mathbf{1}}$}
\psfrag{b}{$\leq\delta^{\mathbf{1.5}}$}
\psfrag{c}{$< \delta^{\mathbf{2}}$}
\psfrag{d}{$\delta^{\mathbf{2}}$}
\psfrag{e}{$<\delta^{\mathbf{2.5}}$}
\psfrag{f}{$<\delta^{\mathbf{3}}$}
\psfrag{g}{$\delta^{\mathbf{3}}$}
\psfrag{h}{$etc$}
\psfrag{i}{$r_1$}
\psfrag{j}{$r_1 = \frac{R}{2}$}
\psfrag{n}{$\tilde{r}_1$}
\psfrag{m}{$r_2$}
\psfrag{l}{$\tilde{r}_1$}
\psfrag{k}{$\tilde{r}_1 = r_2 = \frac{R}{2}$}
\psfrag{o}{$\tilde{r}_1 = \tilde{r}_2$}
\psfrag{p}{$\tilde{r}_1 = \tilde{r}_2$}
\psfrag{r}{$r_3$}
\centerline{\includegraphics{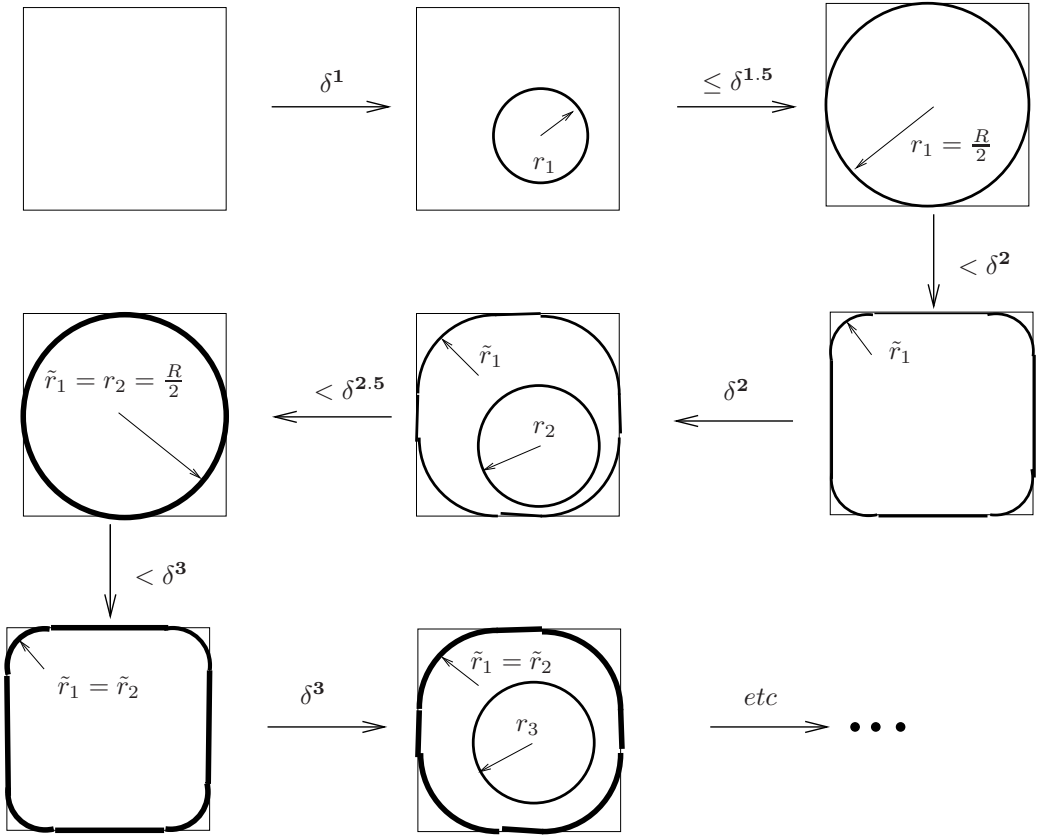}}
\label{Regimes}
\caption{Creation and evolution of macroscopic monolayers on $\Gamma$ as $\delta$ 
grows}
\end{figure}

A complete investigation of the restricted Wulff variational problem 
(see \eqref{12} below) and, accordingly, a rigorous treatment 
of the interface repulsion phenomenon which shows up on the microscopic 
level in all the regimes from \textbf{II.5} on is relegated to a forthcoming 
paper \cite{IS}. 
For the rest of the  paper we shall focus on the regimes \textbf{0, I} 
and \textbf{II} in the context of a simplified model which we 
proceed to introduce. 
  
\section{Our model} 
 
We consider the following lattice model of two-phase coexistence. The 3D box 
\[ 
B_{N}=\Lambda_{N}\times\left\{  -N-1/2,-N+1/2,...,N-1/2,N+1/2\right\} 
\] 
is filled with two kinds of particles: $v$-particles (vapour phase) and 
$s$-particles (solid phase). Here $\Lambda_{N}$ is a two-dimensional $RN\times 
RN$ box; 
\[ 
\Lambda_{N}\,=\,\left\{  0,1,\dots,RN-1\right\}  ^{2}, 
\] 
and 
$R$ is a constant, which we shall set later on to be big enough, in order to 
make our picture reacher. We have $\left\vert B_{N}\right\vert =2R^{2}N^{3}.$ 
Vapour $v$-particles are occupying the upper part of $B_{N},$ while solid 
$s$-particles -- the lower part. Some sites of the box $B_{N}$ can be empty. 
In our model the two phases are separated by an interface $\Gamma$, which is 
supposed to be an SOS-type surface; it is uniquely defined by a function 
\[ 
h_{\Gamma}:\Lambda_{N}^{\circ}\rightarrow\left\{  -N,-N+1,...,N\right\}  , 
\] 
where $\Lambda_{N}^{\circ}$ is the interior of $\Lambda_{N}$. We assume that 
the interface $\Gamma$ is pinned at zero height on the boundary $\partial 
\Lambda_{N}$, that is $h_{\Gamma}\equiv0$ on $\partial\Lambda_{N}$. 
 
Such a surface $\Gamma$ splits $B_{N}$ into two parts; let us denote by $V_{N} 
\left(  \Gamma\right)  $ and $S_{N}\left(  \Gamma\right)  $ the upper and the 
lower halves. The set of configurations of our model consists thus from a 
surface $\Gamma$ plus a choice of two subsets, $\sigma_{v}\subset V_{N} 
\left(  \Gamma\right)  $ and $\sigma_{s}\subset S_{N} \left(  \Gamma\right) 
;$ we have a vapour particle at a point $x\in B_{N}$ iff $x\in\sigma_{v},$ and 
similarly for solid particles. 
 
The partition function $Z_{N}\left(  \beta\right)  $ of our model is now given 
by%
\begin{align} 
&  Z_{N}\left(  \beta\right)  =\nonumber\\ 
&  \sum_{\left(  \Gamma,\sigma_{v},\sigma_{s}\right)  }\exp\left\{ 
-\beta\left\vert \Gamma\right\vert -\left(  a\left\vert \sigma_{v}\right\vert 
+b\left\vert V_{N}\left(  \Gamma\right)  \setminus\sigma_{v}\right\vert 
+c\left\vert \sigma_{s}\right\vert +d\left\vert S_{N} \left(  \Gamma\right) 
\setminus\sigma_{s}\right\vert \right)  \right\}  . \label{02}%
\end{align} 
Here $\left\vert \Gamma\right\vert $ is the surface area of $\Gamma,$ 
$\left\vert \sigma_{v}\right\vert $ is the number of vapour particles, ..., 
while $a,b,c,d$ are four chemical potentials. We want the two phases to be in 
the equilibrium, so we suppose that%
\[ 
e^{-a}+e^{-b}=e^{-c}+e^{-d}\equiv e^{-f}, 
\] 
where the last equality is our definition of the free energy $f.$ Accordingly, 
let us define microscopic occupation probabilities in vapour and solid states 
as 
\[ 
p^{v}\, =\, e^{f -a}\quad\text{and}\quad p^{s}\, =\, e^{f -c}. 
\] 
To mimic the fact that the density of the solid state has to be higher, we 
impose the relation $p^{v} <p^{s}$. 
 
We will study our model under the condition that the total number of particles 
is fixed, and in the leading order of $N$ it is $2\rho R^{2}N^{3},$ with 
$\rho$ between the values $p^{v}$ and $p^{s}$. Of course, flat interface at 
level zero should correspond to the choice 
\[ 
\rho_{0}\, =\, \frac{p^{s} +p^{v}}{2} . 
\] 
More generally, given $\rho= \rho_{0} +\Delta$, one expects to find $\Gamma$ 
to be located approximately at the height $\ell N$ above zero level, where 
$\ell$ satisfies%
\[ 
\frac{\ell}{2}\left(  p^{s} -p^{v}\right)  \, =\, \Delta. 
\] 
The above reasoning suggests that in our model 
the formation of macroscopic monolayers 
over flat interface should 
happen in canonical ensemble with total number of particles being fixed at 
\begin{equation} 
\label{anot}2\rho_{0}R^{2}N^{3} +\delta N^{2}\, \overset{\Delta}{=}\, a_{0} 
N^{3} + \delta N^{2}%
\end{equation} 
with varying $\delta$. 
 
We will denote by $\mathbb{P}$ the (\textquotedblleft grand 
canonical\textquotedblright) probability distribution on triples $\left\{ 
\Gamma,\sigma_{v},\sigma_{s}\right\}  $ corresponding to the above partition 
function. Our main interest in this paper is the study of the conditional 
distribution of the random surface $\Gamma,$ under condition that the total 
\textit{number of particles} 
\[ 
\Sigma\, \overset{\Delta}{=}\, \left\vert \sigma_{v}\right\vert +\left\vert 
\sigma_{s}\right\vert \,\overset{\Delta}{=}\, \Sigma_{v} +\Sigma_{s} , 
\] 
is fixed, i.e. the distribution $\mathbb{P}\left(  \Gamma\, \Bigm|\, \Sigma\, 
=\, a_{0} N^{3} +\delta N^{2} \right)  $. 
 
To study this conditional distribution we rely on Bayes' rule, 
\[ 
\mathbb{P}\left(  \Gamma\Bigm|\Sigma\,=\,a_{0}N^{3}+\delta N^{2}\right) 
\,=\,\frac{\mathbb{P}\left(  \Sigma\,=\,a_{0}N^{3}+\delta N^{2}\Bigm|\Gamma 
\right)  \mathbb{P}\left(  \Gamma\right)  }{\sum_{\Gamma^{\prime}}%
\mathbb{P}\left(  \Sigma\,=\,a_{0}N^{3}+\delta N^{2}\Bigm|\Gamma^{\prime 
}\right)  \mathbb{P}\left(  \Gamma^{\prime}\right)  }. 
\] 
The control over the conditional probabilities $\mathbb{P}\left( 
\bullet\Bigm|\Gamma\right)  $ comes from volume order local limit theorems for 
independent Bernoulli variables, whereas a-priori probabilities $\mathbb{P}%
\left(  \Gamma\right)  $ are derived from representation of $\Gamma$ in terms 
of a gas of non-interacting contours. 
 
In the sequel $c_{1}, c_{2}, \dots$ are positive constants which appear in 
various inequalities and whose values are fixed in such a way that the 
corresponding bounds hold true. 
 
\section{Volume order limit theorems} 
 
The study of probabilities $\mathbf{\Pr}\left(  \Sigma\, =\, a_{0} N^{3} 
+\delta N^{2} \Bigm|\Gamma\right)  $ is easy, since we are dealing with 
independent variables. Indeed, let $B_{N} =S_{N}\cup V_{N}$ be the 
decomposition of $B_{N}$ induced by $\Gamma$. Then, the $\mathbb{ P}\left( 
\bullet\Bigm|\Gamma\right)  $-conditional distribution of the overall number 
of particles is 
\[ 
\Sigma\, =\, \sum_{i\in S_{N}}\xi_{i}^{s}\, +\, \sum_{j\in V_{N}}\xi_{j}^{v}, 
\] 
with iid Bernoulli$(p^{s} )$ random variables $\xi_{i}^{s}$, and iid 
Bernoulli$(p^{v} )$ random variables $\xi_{j}^{v}$. 
 
Let $\alpha(\Gamma)$ be the signed volume under the interface $\Gamma$, 
\begin{equation} 
\alpha(\Gamma)\,=\,\int\int h_{\Gamma}(x,y)dxdy, \label{alpha}%
\end{equation} 
where we set $h_{\Gamma}$ to be equal to $h_{\Gamma}(i)$ in the unit box 
$i+[1/2,1/2]^{2}$. Clearly, $|S_{N}|=R^{2}N^{3}+\alpha(\Gamma)$ and 
$|V_{N}|=R^{2}N^{3}-\alpha(\Gamma)$. Accordingly, 
\[ 
\mathbb{E}\left(  \,\Sigma\,\Bigm|\Gamma\right)  \,=\,a_{0}N^{3}+\alpha 
(\Gamma)p^{sv}, 
\] 
where $p^{sv}\overset{\Delta}{=}p^{s}-p^{v}$. Introducing the variances 
$D^{s}=p^{s}(1-p^{s})$, $D^{v}=p^{v}(1-p^{v})$ and $D=D^{s}+D^{v}$, we infer 
from the Local Limit Theorem (LLT) behavior: For every $K$ fixed there exist 
two positive constants $c_{1}$ and $c_{2}$, such that 
\begin{equation} 
c_{1}\,\leq\,\frac{\mathbb{P}\left(  \Sigma\,=\,a_{0}N^{3}+\delta 
N^{2}\Bigm|\Gamma\right)  }{\frac{1}{\sqrt{\pi D|B_{N}|}}\exp\left\{ 
-\frac{(\alpha(\Gamma)p^{sv}-\delta N^{2})^{2}}{D|B_{N}|}\right\}  }\leq 
c_{2}, \label{10}%
\end{equation} 
uniformly in $N$, $|\delta|\leq K$ and $\Gamma$, provided $|\alpha 
(\Gamma)|\leq KN^{2}$. 
 
\section{Surface weights} 
 
We now want to describe the a-priori probability distribution $\mathbb{P}%
\left(  \Gamma\right)  .$ It is convenient and natural to express it via the 
weights $\left\{  w\left(  \Gamma\right)  \right\}  $, so that 
\begin{equation} 
\mathbb{P}\left(  \Gamma\right)  \,\overset{\Delta}{=}\,\mathbf{\Pr}\left( 
\Gamma\right)  \,=\,\frac{w\left(  \Gamma\right)  }{\sum_{\Gamma}w\left( 
\Gamma\right)  }, \label{03}%
\end{equation} 
where we shall use an additional symbol $\mathbf{\Pr}$ in order to stress that 
the corresponding probabilities are computed in the contour model we are going 
to introduce now. 
 

For our purposes it is necessary to introduce a contour parameterization of 
the set of all surfaces $\Gamma$. Contours will live on the bonds of the dual 
(two dimensional) box $\Lambda_{N}^{*} = \left\{  1/2, 3/2, \dots,RN - 3/2 
\right\}  ^{2}$, and they are defined as follows: Given an interface $\Gamma$ 
and, accordingly, the height function $h_{\Gamma}$ which, by definition, is 
identically zero outside $\Lambda_{N}^{\circ}$, define the following 
semi-infinite subset $\widehat{\Gamma}$ of $\mathbb{R}^{3}$, 
\[ 
\widehat{\Gamma}\, =\, \bigcup_{\overset{(x,y, k)}{k < h_{\Gamma}(x, y)}}\, 
\left(  (x,y, k) +\widehat{C}\right)  , 
\] 
where $\widehat{C} = [-1/2, 1/2]^{3}$ is the unit cube. The above union is 
over all $(x, y)\in\mathbb{Z}^{2}$ and $k\in1/2 +\mathbb{Z}$. 
 
Consider now the level sets of $\Gamma,$ i.e. the sets 
\[ 
H_{k}\, =\, H_{k}\left(  \widehat\Gamma\right)  \, =\, \left\{  \left( 
x,y\right)  \in\mathbb{R}^{2}:\left(  x,y,k\right)  \in\widehat\Gamma\right\} 
,\ k=-N,\, -N+1,\dots, \, N. 
\] 
We define 
\textit{contours} as the connected components of sets $\partial H_{k}%
$\textit{. } The length $\left\vert {\gamma}\right\vert $ of a contour is 
defined in an obvious way. Since, by construction all contours are closed 
polygons composed of the nearest neighbour bonds of $\Lambda_{N}^{*}$, the 
notions of interiour $\mathrm{int}(\gamma)$ and exteriour $\mathrm{ext}%
(\gamma)$ of a contour $\gamma$ are well defined. A contour ${\gamma}$\textit{ 
}is called a $\oplus$-contour ($\ominus$-contour), if the values of the 
function $h_{\Gamma}$ at the immediate exterior of $\gamma$ are smaller 
(bigger) than those at the immediate interiour of $\gamma$. 

Alternatively, let us orient the bonds of each contours $\gamma\subseteq 
\partial H_{k}$ in such a way that when we traverse $\gamma$ the set $H_{k}$ 
remains to the right. Then $\oplus$-contours are those which are clockwise 
oriented with respect to their interiour, whereas $\ominus$-contours are 
counter-clockwise oriented with respect to their interiour. 
 
Let us say that two oriented contours $\gamma$ and $\gamma^{\prime}$ are 
compatible, $\gamma\sim\gamma^{\prime}$, if 
 
\begin{enumerate} 
\item Either $\mathrm{int} (\gamma)\cap\mathrm{int} (\gamma^{\prime 
})=\emptyset$ or $\mathrm{int} (\gamma)\subseteq\mathrm{int} (\gamma^{\prime 
})$ or $\mathrm{int} (\gamma^{\prime})\subseteq\mathrm{int} (\gamma)$. 
 
\item Whenever $\gamma$ and $\gamma^{\prime}$ share a bond $b$, $b$ has the 
same orientation in both $\gamma$ and $\gamma^{\prime}$. 
\end{enumerate} 
 
A family $\Gamma= \left\{  {\gamma}_{i}\right\}  $ of oriented contours is 
called consistent, if contours of $\Gamma$ are pair-wise compatible. It is 
clear that the interfaces $\Gamma$ are in one-to-one correspondence with 
consistent families of oriented contours. The height function $h_{\Gamma}$ 
could be reconstructed from a consistent family $\Gamma=\left\{ 
\gamma\right\}  $ in the following way: For every contour $\gamma$ the sign 
of $\gamma$, which we denote as $\mathrm{sign} (\gamma)$, 
 could be read from it orientation. Then, 
\[ 
h_{\gamma}(x,y )\, =\, \mathrm{sign}(\gamma) \chi_{\mathrm{int}(\gamma)}(x,y ) 
\quad\text{and}\quad h_{\Gamma}\, =\, \sum_{\gamma\in\Gamma} h_{\gamma} , 
\] 
where $\chi_{A}$ is the indicator function of $A$. 
 
We are finally ready to specify the weights $w (\Gamma)$ which appear in 
\eqref{03}: Let $\Gamma=\left\{  \gamma\right\}  $ be a consistent family of 
oriented (signed) contours, Then, 
\begin{equation} 
\label{Gammaweight}w (\Gamma)\, =\, \mathrm{exp}\left\{  -\beta\, \sum 
_{\gamma\in\Gamma}\left\vert \gamma\right\vert \right\}  . 
\end{equation} 
By definition the weight of the flat interface $w (\Gamma_{0} )=1$. 
 
\section{Estimates in the contour ensemble} 
 
In order to make the contour model \eqref{03} , \eqref{Gammaweight} tractable 
one should, evidently, make certain assumptions on the largeness of $\beta$, 
e.g. $e^{\beta}$ should be certainly larger than the connective constant of 
self-avoiding random walks on $\mathbb{Z}^{2}$ \cite{MS}. In fact, it would be 
possible to push for optimal results in terms of the range of $\beta$ along 
the lines of recent developments in the Ornstein-Zernike theory \cite{I, CIV1, 
CIV2}. However, in order to facilitate the exposition and in order to focus on 
the phenomenon of monolayer creation per se, we shall just conveniently assume 
that $\beta$ is so large that one or another form of cluster expansion goes 
through, see eg. \cite{D2}. Due to the ($\pm$-contour) symmetry of the model 
the corresponding techniques would be quite similar to those developed in the 
context of the 2D low temperature Ising model in \cite{DKS}. Consequently, 
instead of stating conditions on $\beta$ explicitly we shall just assume that 
$\beta>\beta_{0}$, where $\beta_{0}$ is so large that all the claims 
formulated below are true. 
 
In the sequel we shall employ the following notation: $\mathcal{C}$ for 
clusters of non-compatible contours and $\Phi_{\beta}(\mathcal{C})$ for the 
corresponding cluster weights which shows up in the cluster expansion 
representation of partition functions. \vskip0.1cm 
 
\noindent\emph{Peierls estimate on appearance of $\gamma$. } Given a contour 
$\gamma$ and a consistent family of contours $\Gamma$, let us say that 
$\gamma\,\overset{k}{\in}\,\Gamma,$ if $\gamma$ appears in $\Gamma$ exactly 
$k$ times. Then, 
\begin{equation} 
\Pr\left(  \gamma\,\overset{k}{\in}\,\Gamma\right)  \,\leq\,e^{-k\beta 
|\gamma|}. \label{Prgamma}%
\end{equation} 
Indeed, every $\Gamma$ satisfying $\gamma\,\overset{k}{\in}\,\Gamma$ can be 
decomposed as $\Gamma=\Gamma^{\prime}\cup\gamma\cup\dots\cup\gamma$. 
Therefore, 
\[ 
\Pr\left(  \gamma\,\overset{k}{\in}\,\Gamma\right)  \,\leq\,\frac{\sum 
_{\Gamma^{\prime}}w(\Gamma^{\prime})e^{-k\beta|\gamma|}}{\sum_{\Gamma^{\prime 
}}w(\Gamma^{\prime})}, 
\] 
where the sums are over all consistent families which are compatible with 
$\gamma$, but do not contain it. \vskip0.1cm 
 
\noindent\emph{Fluctuations of $\alpha(\Gamma)$ and absence of intermediate 
contours}. The following rough a-priori statement is a consequence of 
\eqref{Prgamma}: There exist positive $\nu$ such that for every 
$b_0>0$ fixed,  
\begin{equation} 
\Pr\left(  |\alpha(\Gamma)|\,>\,bN^{2}\right)  \,\leq\,c_{3}e^{-\nu N\sqrt{b}%
}, \label{c3nu}%
\end{equation} 
uniformly in $b\geq b_0$ and in $N$ large enough. 
 
In view of \eqref{10} (computed with respect to the flat interface $\Gamma 
_{0}$ with $\alpha(\Gamma_{0})=0$) the bound \eqref{c3nu} implies that the 
canonical distribution $\mathbb{P}\left(  \,\bullet\,\Big\vert\,\Sigma 
=a_{0}N^{3}+\delta N^{2}\right)  $ is concentrated on $\Gamma$ with 
\begin{equation} 
\alpha(\Gamma)\,\leq\,N^{2}\max\left\{  \frac{\delta^{4}}{\nu^{2}D^{2}R^{4}%
},b_{0}\right\}  . \label{Area}%
\end{equation} 
Now let the interface $\Gamma$ be given by a consistent collection of 
contours, and assume that $\gamma\sim\Gamma$. Of course $\alpha(\Gamma 
\cup\gamma)=\alpha(\Gamma)+\alpha(\gamma)$. Let us assume that the surface 
$\Gamma$ satisfies the estimate \eqref{Area}. Then 
\begin{align*} 
&  \mathbb{P}\left(  \Gamma\cup\gamma\,\Big\vert\,\Sigma=a_{0}N^{3}+\delta 
N^{2}\right)  \,\\ 
&  \leq\,\frac{\mathbb{P}\left(  \Sigma=a_{0}N^{3}+\delta N^{2}%
\,\Big\vert\,\Gamma\cup\gamma\right)  }{\mathbb{P}\left(  \Sigma=a_{0}%
N^{3}+\delta N^{2}\,\Big\vert\,\Gamma\right)  }\,\cdot\,\frac{\Pr\left( 
\Gamma\cup\gamma\right)  }{\Pr\left(  \Gamma\right)  }\\ 
&  \leq c_{4}\exp\left\{  \,c_{5}\frac{|\alpha(\gamma)|}{N}-\beta 
|\gamma|\right\}  \leq c_{4}\exp\left\{  \,c_{6}\frac{|\gamma|^{2}}{N}%
-\beta|\gamma|\right\}  \newline%
\end{align*} 
where we have successively relied on Bayes' rule, \eqref{10} and on the 
isoperimetric inequality. 
 
It follows that for every $K$ there exists $\epsilon=\epsilon(\beta)>0$ such 
that intermediate contours $\gamma$ with 
\begin{equation} 
\frac{1}{\epsilon}\log N<|\gamma|<\epsilon N \label{intermediate}%
\end{equation} 
are, uniformly in $|\delta|<K$, improbable under the conditional distribution 
\[
\mathbb{P}\left(  \,\bullet\,\Big\vert\Sigma=a_{0}N^{3}+\delta N^{2}\right) .
\]
 In the sequel we shall frequently \textit{ignore} intermediate contours, as 
if they do not contribute at all to the distribution \eqref{03}. To avoid 
confusion, we shall use $\widehat{\Pr}$ for the restricted contour ensemble, 
which is defined exactly as in \eqref{03}, except that the intermediate 
contours $\gamma$ satisfying \eqref{intermediate} are suppressed. 
 
\section{The surface tension and the Wulff shape} 
 
Since we are anticipating formation of a monolayer droplet on the interface, 
we are going to need the surface tension function in order to study such a 
droplet and to determine its shape. It is defined in the following way: Let 
$\lambda$ be an oriented site self avoiding path on the dual lattice 
$\mathbb{Z}_{\ast}^{2}$. An oriented contour $\gamma$ is said to be compatible 
with $\lambda$; $\gamma\sim\lambda$, if $\lambda\cap\mathrm{int}%
(\gamma)=\emptyset$ and if whenever $\lambda$ and $\gamma$ share a bond $b$, 
the orientation of $b$ is the same in both $\lambda$ and $\gamma$. 
Accordingly, if $\mathcal{C}$ is a cluster of (incompatible) contours, then 
$\mathcal{C}\sim\lambda$ if $\gamma\sim\lambda$ for every $\gamma 
\in\mathcal{C}$. 
 
In the sequel $0^{\ast}=(1/2,1/2)$ denotes the origin of $\mathbb{Z}_{\ast 
}^{2}$. Let $x\in\mathbb{Z}_{\ast}^{2}$. Set, 
\[ 
T_{\beta}(x)\,=\,\sum_{\lambda:0^{\ast}\rightarrow x}exp\left\{ 
-\beta|\lambda|\,-\,\sum_{\mathcal{C}\not \sim \lambda}\Phi_{\beta 
}(\mathcal{C})\right\}  , 
\] 
where the sum is over all oriented self-avoiding paths from $0^{\ast}$ to $x$. 
 
Let $n\in\mathbb{S}^{1}$ be a unit vector, and $n^{\bot}\in\mathbb{S}^{1}$ is 
orthogonal to it. The surface tension $\tau_{\beta}$ in direction $n\ $is 
defined as 
\[ 
\tau_{\beta}(n)\,=\,-\lim_{L\rightarrow\infty}\frac{1}{L}\log T_{\beta 
}(\lfloor Ln^{\bot}\rfloor). 
\]

Consider the Wulff variational problem, which is a question of finding the 
minimum $w_{\beta}\left(  S\right)  $ of the functional,%
\[ 
w_{\beta}\left(  S\right)  \equiv\min_{\left\{  \lambda:\mathrm{Area}\left( 
\lambda\right)  =S\right\}  }\mathcal{W}\left(  \lambda\right)  . 
\] 
Here 
\[ 
\mathcal{W}\left(  \lambda\right)  =\int_{\lambda}\tau_{\beta}\left( 
n_{s}\right)  ds, 
\] 
$n_{s}$ being the unit normal to $\lambda$ at the point $\lambda\left( 
s\right)  ,$ and the minimum is taken over all closed self-avoiding loops 
$\lambda,$ enclosing the area $S.$ Of course, $w_{\beta}\left(  S\right) 
=\sqrt{S}w_{\beta}\left(  1\right)  .$ Let us denote by $W_{\beta}$ the Wulff 
shape, which is the minimizing loop with area $S=1.$ 
 
As in \cite{DKS} it could be shown that if $\beta$ is sufficiently large, then 
$\tau_{\beta}$ is well defined and strictly positive. Furthermore, the 
boundary of the optimal loop $W_{\beta}$ is locally analytic and has uniformly 
positive and bounded curvature. 
 
One can now apply to the present setting the machinery and the results of 
\cite{DKS}, \cite{DS1}, \cite{DS2}, \cite{SchS} and \cite{ISch}. They allow us 
to study the probabilities of the events 
\begin{equation} 
\mathbf{\Pr}\left(  A_{b}\right)  \equiv\mathbf{\Pr}\left\{  \Gamma 
:\alpha(\Gamma)\,=\,b\right\}  , \label{05}%
\end{equation} 
where we consider here the probability distribution \eqref{03}. 
 
As it 
follows from local limit results in the restricted phase \cite{DKS}
 without intermediate contours \eqref{intermediate}, 
for all values of $b$, 
the probability $\widehat{\mathbf{\Pr}}\left(  A_{b}\right)  $ is bounded 
above by 
\begin{equation} 
\widehat{\mathbf{\Pr}}\left(  A_{b}\right)  \,\leq\,c_{7}\exp\left\{ 
-c_{8}\frac{b^{2}}{N^{2}}\wedge N\right\}  . \label{06}%
\end{equation} 
In particular, for the values of $b\ll N^{3/2}$ the main contribution to 
$\widehat{\mathbf{\Pr}}\left(  A_{b}\right)  $ comes from small contours; 
$|\gamma|<\epsilon^{-1}\log N$. In other words, for such values of $b$, 
conditional distribution $\widehat{\mathbf{\Pr}}\left(  \cdot\Bigm|A_{b}%
\right)  $ is concentrated on the interfaces $\Gamma$ which are essentially 
flat: all contours $\gamma$ of a typical surface $\Gamma$ are less than 
$\epsilon^{-1}\log N$ in length, while their density goes to zero as 
$\beta\rightarrow\infty.$ 
 
On the other hand, for values of $b\gg N^{3/2}$ long contours contribute, and 
the probabilities $\mathbf{\Pr}\left(  A_{b}\right)  $ satisfy%
\begin{equation} 
\log\mathbf{\Pr}\left(  A_{b}\right)  \,=\,-\sqrt{b}w_{\beta}\left(  1\right) 
\left(  1+o_{N}\left(  1\right)  \right)  , \label{07}%
\end{equation} 
provided, of course, that the scaled Wulff shape $\sqrt{b/N^{2}}~W_{\beta}$ 
fits into the square $[0,R]^{2}$. 
Under these two restrictions on $b$ the analysis of \cite{DKS} implies that 
the conditional distribution $\mathbf{\Pr}\left(  \cdot\Bigm|A_{b}\right)  $ 
is concentrated on the interfaces $\Gamma$ which are \textquotedblleft 
occupying two consecutive levels\textquotedblright. Namely, the set $\left\{ 
\gamma_{i}\right\}  $ of contours, comprising $\Gamma,$ contains exactly one 
large contour, $\gamma_{0},$ of diameter $\sim\sqrt{b},$ while the rest of 
them have their lengths not exceeding $\epsilon^{-1}\ln N$. 
The contour $\gamma_{0}$ is of $\oplus$-type, so for the majority of points 
inside $\gamma_{0}$ the value of the height function $h_{\Gamma}$ is $1,$ 
while outside $\gamma_{0}$ it is mainly zero. Finally, the contour $\gamma 
_{0}$ has 
 
\begin{itemize} 
\item \textit{Asymptotic shape}: The contour $\gamma_{0}$ 
is of size $\sim\sqrt{b},$ and it follows very close the curve $\sqrt 
{b}W_{\beta}.$ Namely, the latter can be shifted in such a way that the 
Hausdorff distance 
\begin{equation} 
\rho_{H}\left(  {\gamma}_{0},\sqrt{b}W_{\beta}\right)  \leq\sqrt[3]{b}. 
\label{13}%
\end{equation} 
 
\end{itemize} 
 
Of course, all the claims above should be understood to hold only on the set 
of typical configurations, i.e. on the sets of (conditional) probabilities 
going to $1$ as $N\rightarrow\infty.$ 
 
In the present paper we also  
we need to consider such values of $b\sim2R^{2}N^{2},$ when  
the scaled Wulff shape $\sqrt{b/N^{2}}%
~W_{\beta}$ does not fit into the square $[0,R]^{2}$. This situation was 
partially treated in the paper \cite{SchS}, and the technique of that paper 
provides us with the following information about the typical behavior of 
$\Gamma$ under the distribution $\widehat{\mathbf{\Pr}}\left(  \cdot 
\Bigm|A_{b}\right)  $ for the remaining values of $b.$ 
 
Namely, instead of the Wulff variational problem we have to consider the 
following \textit{restricted }Wulff variational problem, which is a problem of 
finding the minimum 
\begin{equation} 
w_{\beta}^{rst}\left(  S\right)  \equiv\min_{\left\{  k;\lambda_{1}%
,...,\lambda_{k}\right\}  }\mathcal{W}_{S}^{rst}\left(  k;\lambda 
_{1},...,\lambda_{k}\right)  \equiv\mathcal{W}\left(  \lambda_{1}\right) 
+...+\mathcal{W}\left(  \lambda_{k}\right)  , \label{12}%
\end{equation} 
where 
 
\begin{itemize} 
\item the curves $\lambda_{1},...,\lambda_{k}$ are closed piecewise smooth 
loops \textit{inside }the unit square $Q_{1}$; 
 
\item the loops $\lambda_{i}$ are nested: $\mathrm{Int}\left(  \lambda 
_{k}\right)  \subseteq\mathrm{Int}\left(  \lambda_{k-1}\right)  \subseteq 
...\subseteq \mathrm{Int}\left(  \lambda_{1}\right)  ;$ 
 
\item $\mathrm{Area}\left(  \lambda_{k}\right)  +\mathrm{Area}\left( 
\lambda_{k-1}\right)  +...+\mathrm{Area}\left(  \lambda_{1}\right)  =S.$ 
\end{itemize} 
 
\noindent The parameter $k$ is not fixed; we have to minimize over $k$ as 
well. For the area parameter $S$ small enough, the minimum in $\left( 
\ref{12}\right)  $ is attained at $k=1,$ while $\lambda_{1}$ is the scaled 
Wulff shape, $\sqrt{S}W_{\beta}.$ In other words, in this regime $w_{\beta 
}^{rst}\left(  S\right)  =w_{\beta}\left(  S\right)  .$ Let $S_{1}$ be the 
maximal value, for which the inclusion $\sqrt{S}W_{\beta}\subset Q_{1}$ is 
possible. In the range $S_{1}<S<1$ the solution to $\left(  \ref{12}\right)  $ 
is given by $k=1,$ while the loop $\lambda_{1}$ is the corresponding Wulff 
plaquette, described above. In the range $1<S<2S_{1}$ the solution has the 
value $k=2,$ the curve $\lambda_{1}$ is the Wulff plaquette, while the curve 
$\lambda_{2}\subset\lambda_{1}$ is the Wulff shape; they are uniquely defined 
by the two conditions: 
 
\begin{enumerate} 
\item $\mathrm{Area}\left(  \lambda_{2}\right)  +\mathrm{Area}\left( 
\lambda_{1}\right)  =S,$ 
 
\item the curved parts of $\lambda_{1}$ are translations of the corresponding 
quarters of $\lambda_{2}.$ 
\end{enumerate} 
 
\noindent In the range $2S_{1}<S<2$ we have $k=2,$ while the loops 
$\lambda_{2}=\lambda_{1}$ are identical Wulff plaquettes. 
 
The relation $\left(  \ref{07}\right)  $ is generalized to 
\begin{equation} 
\log\mathbf{\Pr}\left(  A_{b}\right)  \,=\,-RNw_{\beta}^{rst}\left(  \frac 
{b}{R^{2}N^{2}}\right)  \left(  1+o_{N}\left(  1\right)  \right)  . \label{20}%
\end{equation}

The function $w_{\beta}^{rst}\left(  S\right)  $ is evidently increasing in 
$S.$ For $S$ small it behaves as $c^{\prime}\sqrt{S}.$ In the vicinity of the 
point $S=1$ it behaves as $c^{\prime\prime}\sqrt{S-1}$ for $S>1,$ and as 
$c^{\prime\prime\prime}\sqrt{1-S}$ for $S<1.$ Otherwise it is a smooth 
function of $S,$ $0\leq S<2.$ The two singularities we just pointed out, are 
responsible for the interesting geometric behavior of our model, which has 
been described informally in Section 2, and will be explicitely formulated in 
the next Section. Namely, each one is responsible for the appearance of the 
corresponding droplet. 
 
Accordingly, once the Wulff shape $\sqrt{b/N^{2}}~W_{\beta}$ does not fit into 
the square $[0,R]^{2}$, while $b\leq 
c\left(  \beta\right) 
R^{2}N^{2}$ ( where the constant 
 $c\left(  \beta\right)  \rightarrow1$ {as }$\beta 
\rightarrow\infty$)  the conditional distribution $\mathbf{\Pr 
}\left(  \cdot\Bigm|A_{b}\right)  $ is concentrated on the interfaces $\Gamma$ 
which again are occupying two consecutive levels. The set $\left\{  \gamma 
_{i}\right\}  $ of contours, comprising $\Gamma,$ contains one large contour, 
$\gamma_{0},$ this time of diameter $\sim R,$ which in some places is going 
very close to the boundary of our box. The rest of contours have their lengths 
not exceeding $\epsilon^{-1}\ln N$. The contour $\gamma_{0}$ is of $\oplus 
$-type, and for the majority of points inside $\gamma_{0}$ the value of the 
height function $h_{\Gamma}$ is $1,$ while outside $\gamma_{0}$ it is mainly 
zero. Finally, the contour $\gamma_{0}$ has asymptotic shape of the Wulff 
plaquette, in the same sense as in $\left(  \ref{13}\right)  .$ 
 
In the remaining range $R^{2}N^{2}\leq b\leq2R^{2}N^{2}$ the set $\left\{ 
\gamma_{i}\right\}  $ of contours, comprising $\Gamma,$ contains exactly two 
large contours, $\gamma_{0}$ and $\gamma_{1},$ with $\gamma_{1}\subset 
\mathrm{Int}\left(  \gamma_{0}\right)  ,$ both of the $\oplus$-type. The 
interface $\Gamma$ is, naturally, occupying three consecutive levels: it is 
(typically) at the height $2$ inside $\gamma_{1},$ at height $1$ between 
$\gamma_{0}$ and $\gamma_{1},$ and at height $0$ outside $\gamma_{0}.$ Note 
that for $b$ close to $R^{2}N^{2}$ the contour $\gamma_{1}$ is free to move 
inside $\gamma_{0},$ so its location is random (as is also the case in the 
regime of the unique large contour, when the scaled Wulff shape $\sqrt 
{b/N^{2}}~W_{\beta}$ fits into the square $[0,R]^{2}$). The contour 
$\gamma_{0},$ on the other hand, is (nearly) touching all four sides of the 
boundary of our box, so it is relatively less free to fluctuate. 
 
In the complementary regime, when $b$ is close to $2R^{2}N^{2},$ the two 
contours $\gamma_{0}$ and $\gamma_{1}$ have the same size in the leading order 
(which is linear in $N$), while the Hausdorff distance between them is only 
$\sim N^{1/2};$ it is created as a result of the entropic repulsion between 
them. In particular, in the limit as $N\rightarrow\infty,$ and under the 
$\frac{1}{N}$ scaling, the two contours coincide, going in asymptotic shape to 
the same Wulff plaquette. The study of this case needs the technique, 
additional to that contained in \cite{DKS}, \cite{DS1}, \cite{DS2}, 
\cite{SchS} and \cite{ISch}, since the case of two repelling large contours 
was not considered there. The case of the values $b$ above $2R^{2}N^{2}$ is 
even more involved, since there we have to deal with several large mutually 
repelling contours. We will return to it in a separate publication, see 
\cite{IS}. 
 
\section{Main result} 
 
We are ready now to describe the monolayers creation in our model: Let us fix 
$p^{s}>p^{v}$ (and hence $p^{sv}$ and $D$), and let $\beta$ be sufficiently 
large. Let us also fix $R$ large enough, so that the rescaled Wulff shape of 
area 
\[ 
\sqrt[3]{\frac{D^{2}w_{\beta}^{2}\left(  1\right)  }{p^{sv}}}\,R^{4/3}%
\] 
fits into the $R\times R$ square. 
 
\begin{thm} 
Let $\Gamma$ be a typical interface drawn from the conditional distribution 
$\mathbb{P}\left(  \,\bullet\,\big\vert\,\Sigma=a_{0}N^{3}+\delta 
N^{2}\right)  $. Define 
\begin{equation} 
\delta^{\mathbf{1}}\,=\,\frac{3}{2}\sqrt[3]{D^{2}w_{\beta}^{2}p^{sv}}R^{4/3}. 
\label{deltac}%
\end{equation}

\begin{itemize} 
\item For values of $\delta$ satisfying $0<\delta<\delta^{\mathbf{1}}$, the interface 
$\Gamma$ is essentially flat: all contours of $\Gamma$ have lengths bounded 
above by $\epsilon^{-1}\log N$. 
 
\item There exists $\delta^{\mathbf{2}}> \delta^{\mathbf{1}}$, such that for $\delta^{\mathbf{1}}%
<\delta<\delta^{\mathbf{2}}$ the interface $\Gamma$ has one monolayer. Precisely, 
$\Gamma$ contains exactly one large contour $\gamma_{0}$ of approximately 
Wulff shape (or Wulff plaquette shape), such that 
\begin{equation} 
\alpha(\gamma_{0})\,>\,\frac{2\delta}{3p^{sv}}\,N^{2}. \label{gamma0}%
\end{equation} 
The rest of contours of $\Gamma$ are small; their lengths are bounded above by 
$\epsilon^{-1}\log N$. 
 
\item Similarly, there exists a value $\delta_{R},$ such that for 
$\delta^{\mathbf{2}}<\delta<\delta_{R}$ the interface $\Gamma$ has two monolayers, and 
contains exactly two large contours, $\gamma_{0}$ and $\gamma_{1}%
\subset\mathrm{Int}\left(  \gamma_{0}\right)  $. The bigger one, $\gamma_{0},$ 
has the shape of the Wulff plaquette, while the smaller one has the Wulff 
shape. Again, $\alpha(\gamma_{1})\,>\,\frac{2\delta}{3p^{sv}}\,N^{2}.$ 
\end{itemize} 
\end{thm} 
 
\section{Proof of the main result} 
 
Let us fix $\delta$ and consider the surface distribution $\mathbb{P}\left( 
\,\bullet\,\big\vert\,\Sigma=a_{0}+\delta N^{2}\right)  $. Since we can ignore 
intermediate contours \eqref{intermediate} and since we already know how the 
typical surfaces looks like in the constraint ensembles $\widehat{\Pr}\left( 
\,\bullet\,\big\vert\,A_{b}\right)  $, it would be enough to study conditional 
probabilities $\mathbb{P}\left(  \,A_{b}\,\big\vert\,\Sigma=a_{0}+\delta 
N^{2}\right)  $. Namely, for every $\delta$ we need to know the range of the 
typical values of the \textquotedblleft volume\textquotedblright\ observable 
$b.$ To do this we will compare the probabilities $\mathbb{P}\left( 
\,A_{b}\,,\,\Sigma=a_{0}+\delta N^{2}\right)  \sim\mathbb{P}\left( 
\,\,\Sigma=a_{0}+\delta N^{2}\big\vert\,A_{b}\right)  \widehat{\Pr}\left( 
A_{b}\right)  $ for various values of $b,$ in order to find the dominant one. 
 
There are three regimes to be worked out: Fix $\eta\in\ (0,1/2)$ and $c_{9}$ 
small enough. \vskip0.2cm 
 
\noindent C{\small ase}~1. $b\leq N^{1+\eta}$. By \eqref{10} and \eqref{06}, 
\begin{equation}%
\begin{split} 
&  c_{10}exp\left\{  -\frac{\delta^{2}}{2DR^{2}}N-O\left(  \frac{b^{2}}{N^{2}%
}\right)  \right\} \\ 
&  \leq\,N^{3/2}\mathbb{P}\left(  \Sigma=a_{0}+\delta N^{2}\,\big\vert\,A_{b}%
\right)  \widehat{\Pr}\left(  A_{b}\right) \\ 
&  \leq\,c_{11}exp\left\{  -\frac{\delta^{2}}{2DR^{2}}N\right\}  . 
\end{split} 
\label{case1}%
\end{equation} 
\vskip0.2cm 
 
\noindent C{\small ase}~2. $N^{1+\eta}<b\leq c_{9}N^{2}$. By \eqref{10} and 
\eqref{06}, 
\begin{equation} 
\mathbb{P}\left(  \Sigma=a_{0}+\delta N^{2}\,\big\vert\,A_{b}\right) 
\widehat{\Pr}\left(  A_{b}\right)  \,\leq\,c_{12}exp\left\{  -\frac{\delta 
^{2}}{2DR^{2}}N+\frac{\delta p^{sv}b}{NR^{2}D}-c_{8}\frac{b^{2}}{N^{2}}\wedge 
N\right\}  . \label{case2}%
\end{equation} 
Obviously, once $c_{9}$ is chosen to be sufficiently small, the right hand 
side of \eqref{case2} is negligible with respect to the lower bound on 
left-hand side of \eqref{case1} (computed at $b\ll N^{1+\eta}$). \vskip0.2cm 
 
\noindent C{\small ase}~3. $b=\rho N^{2}$ with $\rho>c_{9}$. By $\left( 
\ref{20}\right)  $ and, once again, by volume order local limit result 
\eqref{10}, 
\begin{equation}%
\begin{split} 
&  exp\left\{  -\frac{(\delta-p^{sv}\rho)^{2}}{DR^{2}}N-RNw_{\beta}%
^{rst}\left(  \frac{\rho}{R^{2}}\right)  -o(N)\right\} \\ 
&  \leq\,\mathbb{P}\left(  \Sigma=a_{0}+\delta N^{2}\,\big\vert\,A_{b}\right) 
\Pr\left(  A_{b}\right) \\ 
&  \leq\,exp\left\{  -\frac{(\delta-p^{sv}\rho)^{2}}{2DR^{2}}N-RNw_{\beta 
}^{rst}\left(  \frac{\rho}{R^{2}}\right)  +o(N)\right\}  . 
\end{split} 
\label{case3}%
\end{equation}

Therefore, in order to figure out the dominant contribution between 
\eqref{case1} and \eqref{case3}, we have to find the global minimum of the 
function%
\begin{equation} 
\frac{\left(  \delta-p^{sv}\rho\right)  ^{2}}{2DR^{2}}+Rw_{\beta}^{rst}\left( 
\frac{\rho}{R^{2}}\right)  \label{22}%
\end{equation} 
on the interval $\rho\in\lbrack0,2R^{2}]$. This minimization problem needs 
just the elementary calculus, see e.g. \cite{BCK}.
  For small values of $\rho$ our function reduces 
to $\frac{\left(  \delta-p^{sv}\rho\right)  ^{2}}{2DR^{2}}+w_{\beta}\left( 
1\right)  \sqrt{\rho}.$ After the following change of variables: 
\[ 
\lambda\,=\,\frac{p^{sv}\rho}{\delta}\quad\text{and}\quad\kappa\,=\kappa 
\left(  \delta\right)  =\,\frac{\delta^{3/2}}{2DR^{2}w_{\beta}\left( 
1\right)  \sqrt{p^{sv}}}, 
\] 
we have to look for global minimizers of 
\[ 
\phi_{\kappa}(\lambda)\,\overset{\Delta}{=}\,\kappa(1-\lambda)^{2}%
+\sqrt{\lambda}. 
\] 
Set 
\begin{equation} 
\kappa_{c}\,=\kappa\left(  \delta^{\mathbf{1}}\right)  =\,\frac{1}{2}\left(  \frac 
{3}{2}\right)  ^{3/2}.\label{kappac}%
\end{equation} 
One easily sees that 
 
\begin{itemize} 
\item If $\kappa<\kappa_{c}$, then the global minimizer is $0$. 
 
\item If $\kappa=\kappa_{c}$ then there are exactly two global minimizers; $0$ 
and $\lambda_{c} = 2/3$. 
 
\item If $\kappa>\kappa_{c}$, then the global minimizer $\lambda_{m}$ is the 
maximal solution of 
\[ 
4\kappa\sqrt{\lambda}\left(  1-\lambda\right)  , 
\] 
which, in particular, satisfies $\lambda_{m}>2/3$. 
\end{itemize} 
 
A similar analysis applies in the vicinity of the singularity of the function 
$w_{\beta}^{rst}\left(  \frac{\rho}{R^{2}}\right)  $ at $\frac{\rho}{R^{2}%
}\sim1.$ Since the function $w_{\beta}^{rst}\left(  S\right)  $ is monotone, 
and has the derivative equal to $+\infty$ at $S=1,$ the point of the global 
minimum of $\left(  \ref{22}\right)  $, which is a monotone function of 
$\delta,$ never belongs to some neighborhood of the point $\frac{\rho}{R^{2}%
}=1.$ Therefore at some $\delta=\delta^{\mathbf{2}}$ it jumps from some value 
$\rho_{-}<R^{2}$ to $\rho_{+}>R^{2}.$ 
 
The proof of Theorem~1 is, thereby, completed. 
 
\section{Conclusions} 
 
In this paper we have described a model of the interface between the vapour 
and liquid phases, evolving as the total number of particles increases. We 
have shown that the evolution of the interface goes via the spontaneous 
formation on it of one monolayer of the size of the system. We believe that 
the same result can be proven for the 3D Ising model with the same boundary 
conditions, i.e. periodic in two horizontal directions and $\pm$ in the 
vertical one. It will be very interesting to establish the phenomenon of the 
monolayer formation in the 3D Ising model with $\left(  +\right)  $-boundary 
conditions, when the monolayer attaches itself to a facet of the Wulff-like 
(random) crystal. This problem, however, seems to be quite difficult, since 
one needs to control the rounded part of the crystal. This rounded part is 
probably behaving as a massless Gaussian random surface (compare with 
\cite{K}), and this alone indicates enough the complexity of the problem.


\begin{thebibliography}{9999}                                                                                             %
 
 
\bibitem[BCK]{BCK}Biskup, M., Chayes, L. and Kotecky, R.: Critical Region for 
Droplet Formation In the Two-Dimensional Ising Model, Comm. Math. Phys., v. 
242, pp 137-183, 2003. 
 
\bibitem[BSS]{BSS}Bodineau, T., Schonmann, R. and Shlosman, S.: 3D Crystal: 
How Flat its Flat Facets Are? Comm. Math. Phys., v. 255, Number 3, pp 747 - 
766, 2005. 
 
\bibitem[CIV1]{CIV1}M.~Campanino, D.~Ioffe and Y.~Velenik: 
\textit{Ornstein-Zernike theory for finite range Ising models above $T\sb c$}, 
Probab. Theory Related Fields 125 (2003), no. 3, 305--349. 
 
\bibitem[CIV2]{CIV2}M.~Campanino, D.~Ioffe and Y.~Velenik: \textit{ 
Fluctuation theory of connectivities in sub-critical random cluster models}, 
to appear in Annals of Probability (2007). 
 
\bibitem[D1]{D1}R.L.~Dobrushin. \textit{Gibbs states describing a coexistence 
of phases for the three-dimensional Ising model}, Teor. Ver. i ee Primeneija, 
17, 582-600, (1972). 
 
\bibitem[D2]{D2}R.~Dobrushin, P.~Groeneboom and M.~Ledoux: \textit{Lectures on 
probability theory and statistics} Lectures from the 24th Saint-Flour Summer 
School held July 7--23, 1994. Edited by P. Bernard. Lecture Notes in 
Mathematics, 1648. Springer-Verlag, Berlin, (1996). 
 
\bibitem[DKS]{DKS}R.L. Dobrushin, R. Kotecky and S. B. Shlosman: \textit{Wulff 
construction: a global shape from local interaction, }AMS translations series, 
Providence (Rhode Island), 1992. 
 
\bibitem[DS1]{DS1}R. L. Dobrushin and S. Shlosman: \textit{Large and moderate 
deviations in the Ising model}, In: \textquotedblright Probability 
contributions to statistical mechanics\textquotedblright, R. L. Dobrushin ed., 
\textquotedblright Advances in Soviet Mathematics\textquotedblright, v. 18, 
pp.91--220, AMS, Providence, RI, 1994 
 
\bibitem[DS2]{DS2}R. L. Dobrushin and S. Shlosman: \textit{Droplet 
condensation in the Ising model: moderate deviations point of view}, 
Proceedings of the NATO Advanced Study Institute:\textquotedblright 
Probability theory of spatial disorder and phase transition\textquotedblright, 
G. Grimmett ed., Kluwer Academic Publishers, vol. 20, pp. 17--34, 1994 
 
\bibitem[HIK]{HIK}O. Hryniv, D. Ioffe and R. Kotecky, \textit{ in preparation} (2007). 
 
\bibitem[K]{K}R. Kenyon: \textit{Dominos and the Gaussian free field}, Ann. 
Prob. 29, no. 3 (2001), 1128-1137. 
 
\bibitem[I]{I}D.~Ioffe: \textit{Ornstein-Zernike behaviour and analyticity of 
shapes for self-avoiding walks on $Z\sp d$}, Markov Process. Related Fields 4 
(1998), no. 3, 323--350. 
 
\bibitem[ISch]{ISch}D.~Ioffe and R.H.~Schonmann: 
\textit{Dobrushin-Koteck\'{y}-Shlosman theorem up to the critical 
temperature}, Comm. Math. Phys. 199 (1998), no. 1, 117--167. 
 
\bibitem[IS]{IS}D.~Ioffe and S.~Shlosman: \textit{Ising model fog drip, II: 
the puddle. }In preparation. 
 
\bibitem[MS]{MS}N.~Madras and G.~Slade: \textit{The self-avoiding walk}: 
Probability and its Applications. Birkhauser Boston, Inc., Boston, MA, (1993). 
 
\bibitem[SchS]{SchS}R.H.~Schonmann and S.~Shlosman: \textit{Constrained 
variational problem with applications to the Ising model}: J. Statist. Phys. 
83 (1996), no. 5-6, 867--905. 
\end{thebibliography}
\end{document}